\documentclass[usenatbib]{mnras}
\usepackage{amsmath,amssymb}
\usepackage[pdftex]{graphicx}

\def\vb#1{\mbox{\boldmath $#1$}}

\newcommand{\Caffau}{$\rm SDSS \ J102915+172927$}

\newcommand{\percc}{{\rm cm^{-3}}}

\newcommand{\E}[1]{\times 10^{#1}}

\newcommand{\nH}{n_{{\rm H}}}

\newcommand{\nHcen}{n_{\rm H, cen}}

\newcommand{\HII}{H~{\sc ii}}

\newcommand{\Rjeans}{R_{\rm J}}
\newcommand{\MPopIII}{M_{\rm Pop III}}
\newcommand{\Esn}{E_{{\rm SN}}}
\newcommand{\erg}{\rm erg}
\newcommand{\tff}{t_{\rm ff}}

\newcommand{\tlife}{t_{\rm life}}
\newcommand{\trecol}{t_{\rm recol}}
\newcommand{\zrecol}{z_{\rm recol}}
\newcommand{\tturb}{t_{\rm turb}}

\newcommand{\Nngb}{N_{\rm ngb}}

\newcommand{\cs}{c_{\rm s}}

\newcommand{\abM}[1]{M_{\rm {#1}}}
\newcommand{\abAsun}[1]{A_{\bigodot}({\rm {#1}})}

\newcommand{\abA}[1]{A({\rm {#1}})}
\newcommand{\abH}[1]{{\rm [{#1}/H]}}

\newcommand{\abX}[1]{X({\rm {#1}})}
\newcommand{\abFe}[1]{{\rm [{#1}/Fe]}}
\newcommand{\muX}[1]{\mu_{\rm {#1}}}

\newcommand{\XH}{X_{{\rm H}}}

\newcommand{\Msun}{{\rm M_{\bigodot}}}

\newcommand{\Mvir}{M_{\rm vir}}
\newcommand{\Rvir}{R_{\rm vir}}

\newcommand{\Mr}{M_{\rm r}}
\newcommand{\ffb}{f_{\rm enr}}
\newcommand{\Rs}{R_{\rm sh}}
\newcommand{\Rf}{R_{\rm fil}}
\newcommand{\Mcl}{M_{\rm cl}}
\newcommand{\Rcl}{R_{\rm cl}}

\defcitealias{Chiaki15}{C15}
\defcitealias{Chiaki16}{C16}
\defcitealias{Chiaki18}{C18}
\defcitealias{Chiaki19}{C19}
\defcitealias{Smith15}{S15}
\defcitealias{Hirano14}{H14}
\defcitealias{Ishigaki14}{I14}
\defcitealias{Tominaga14}{T14}
\defcitealias{Marassi15}{M15}

\title[Structure of Pop III supernova ejecta]
      {Does the structure of Pop III supernova ejecta affect the elemental abundance of extremely metal-poor stars?}

\author[G. Chiaki \& N. Tominaga]
{Gen Chiaki$^{1}$\thanks{E-mail: gen.chiaki@physics.gatech.edu} and
Nozomu Tominaga$^{2, 3}$
\\
$^{1}$Center for Relativistic Astrophysics, School of Physics, Georgia Institute of Technology, Atlanta, GA 30332, USA \\
$^{2}$Department of Physics, Konan University, 8-9-1 Okamoto, Kobe, 658-0072, Japan \\
$^{3}$Kavli Institute for the Physics and Mathematics of the Universe (WPI), 
The University of Tokyo, \\5-1-5 Kashiwanoha, Kashiwa, Chiba 277-8583, Japan}

\begin{document}

\date{}

\pagerange{\pageref{firstpage}--\pageref{lastpage}} \pubyear{2020}

\maketitle

\label{firstpage}

\begin{abstract}
The first generation of metal-free (Pop III) stars are crucial for the production of heavy elements 
in the earliest phase of structure formation.
Their mass scale can be derived from the elemental abundance pattern of extremely metal-poor (EMP) stars,
which are assumed to inherit the abundances of uniformly mixed supernova (SN) ejecta.
If the expanding ejecta maintains its initial stratified structure, 
the elemental abundance pattern of EMP stars might be different from that from uniform ejecta.
In this work we perform numerical simulations of the metal enrichment from stratified ejecta
for normal core-collapse SNe (CCSNe) with a progenitor mass 
$25 \ \Msun$ and explosion energies 0.7--10 B ($1 \ {\rm B} = 10^{51} \ \erg$).
We find that SN shells fall back into the central minihalo in all models.
In the recollapsing clouds, the abundance ratio $\abFe{M}$ for stratified ejecta 
is different from the one for uniform ejecta only within $\pm 0.4$ dex for any element M.
We also find that, for the largest explosion energy (10 B), a neighboring halo is also enriched.
Only the outer layers containing Ca or lighter elements reach the halo, where $\abFe{C} = 1.49$.
This means that C-enhanced metal-poor (CEMP) stars can form from the CCSN 
even with an average abundance ratio $\abFe{C} = -0.65$.
\end{abstract}


%%%%%%%%%%%%%%%%%%%%%%%%%%%%%%%%%%%%%%%%%%%%%
% INTRODUCTION %%%%%%%%%%%%%%%%%%%%%%%%%%%%%%%%%%%
%%%%%%%%%%%%%%%%%%%%%%%%%%%%%%%%%%%%%%%%%%%%%

\section{INTRODUCTION}

The first generation of metal-free (Population III; Pop III) stars 
enrich the early Universe with the first heavy elements, 
affecting later structure formation.
Pop III stars form in dark matter halos with masses $\sim 10^5$--$10^6 \ \Msun$
(minihalos; MH) mainly at redshifts $\sim 10$--30.
If they are massive ($\geq 8 \ \Msun$), metals ejected from their supernova (SNe) can 
enhance star formation \citep[e.g.,][]{Ritter16}.
The efficiency of their feedback
depends on stellar mass and SN explosion energy,
which in turn change the luminosity and metal ejecta mass, respectively.
Theoretical works suggest that Pop III stars are massive ($\MPopIII \sim 10$--$1000 \ \Msun$)
because gas cooling caused by only molecular hydrogen is inefficient, and
their parent clouds are stable against fragmentation
\citep{Bromm99, Abel02, Hirano14, Susa14}.
Although recent numerical studies have found that low-mass Pop III stars
($\lesssim 1 \ \Msun$) can form through the fragmentation of accretion disks
\citep[e.g.,][]{Clark11, Greif12, Stacy14}, we restrict our focus on
massive stars to study their metal enrichment.

A number of previous works have attempted to constrain the mass of Pop III stars.
One approach is to directly observe Pop III stars, but
no metal-free stars have so far been found albeit the 
large survey programs of stars 
in the Milky Way and neighboring dwarf galaxies, such as the HK \citep{Beers85, Beers92},
Hamburg/ESO \citep{Christlieb03}, SEGUE \citep{Yanny09}, and LAMOST \citep{Cui12, Deng12} surveys.
The lower limit of iron abundance $\abH{Fe}$,\footnote{The number abundance 
ratio of an element A to B is conventionally given with
\begin{equation}
{\rm [A/B]} = (\abA{A} - \abA{B}) - (\abAsun{A} - \abAsun{B}), \nonumber
\end{equation}
where $\abA{A}$ is the absolute abundance defined as
\begin{equation}
\abA{A} = 12 + \log(y_{\rm A}), \nonumber
\end{equation}
and $y_{\rm A}$ is the number fraction of A relative to hydrogen nuclei.}
which is often used as a proxy to the stellar metallicity,
of the stars ever observed is $-4.71$ \citep[\Caffau ;][]{Caffau11} for carbon-normal stars.
Some of the so-called carbon-enhanced metal-poor (CEMP) stars have even smaller fractions of Fe ($\abH{Fe} < -5$),
but they have enhanced carbon abundances 
$\abFe{C} > 0.7$ \citep{Beers05, Aoki07}.
Another approach is to measure the elemental abundance 
ratio of extremely metal-poor  (EMP) stars with metallicities $\abH{Fe} < -3$, which are considered to form
from clouds enriched by a single or several Pop III SNe \citep{Ryan96, Cayrel04}.
\citet{Ishigaki18} showed that the elemental abundance ratios of EMP stars are best fit with the Pop III 
SN models with progenitor masses $\MPopIII < 40 \ \Msun$.

This reverse-engineering assumes
that the expanding ejecta of Pop III SNe is uniformly mixed, and
the averaged elemental abundance ratio in all layers of the ejecta is used 
when fitting the stellar abundances.
According to stellar evolution models, the SN ejecta should be initially stratified,
where heavier elements, such as Fe, are in the inner layers
and lighter elements, such as C, are in the outer layers.
If the expanding ejecta maintains its radial structure, it is expected that the abundance
of star-forming clouds enriched by the SNe and stars that eventually form might be
different from that from uniform ejecta.
Supposing that these stars inherit the same abundance from their parent clouds, 
a bias is expected between the elemental abundances of observed EMP stars and monolithic SN ejecta. 

In this work, we investigate the effect of stratified SN ejecta on the elemental 
abundance ratio of the succeeding generations of stars with cosmological simulations.
In our previous work \citet[][hereafter \citetalias{Chiaki18}]{Chiaki18},
we found that the three-dimensional density and velocity structures around the MHs 
have large effects on the dynamical evolution of SN shells.
We also found that, SN shells fall back into MHs that originally hosted Pop III stars and 
collapse again when the smaller explosion energy $\Esn$ is smaller than the halo binding energy.
If the explosion energy is larger, neighboring MHs can also be enriched by Pop III SNe.
These two distinctive enrichment modes are called internal enrichment (IE) and external enrichment (EE).
The ratio of lighter to heavier elements is expected to be different between in the two cases
where the SN ejecta is stratified and uniform.

The bias of an elemental abundance pattern in a recollapsing cloud from that of uniform ejecta was
studied by \citet{Ritter15} and \citet{Sluder16} in the IE case.
They performed a high-resolution simulation of metal enrichment from a Pop III SN,
following the metal dispersion with tracer particles.
They found that the mass fraction of metals originally in the innermost hotter layers 
escaping from the MH is three times larger relative to the metals in the outer layers.
This would result in an enhancement of $\sim 0.5$ dex the abundance ratio of lighter elements 
to heavier elements.
However, they only considered one explosion energy
$\Esn = 1$ B ($1 \ {\rm B} = 10^{51} \ \erg$).
The shell evolution and the resulting elemental abundance ratio of recollapsing clouds are expected to
depend on $\Esn$.
SNe with higher explosion energies ($\sim 10$ B; hypernovae) have been observed,
accompanied with long-duration gamma-ray bursts \citep{Iwamoto98}. 
The elemental abundances of some EMP stars are best fit with hypernova moedels \citep{Tominaga14}.
This motivates us to consider higher explosion energies.
In addition, 
they did not consider the radial distribution of individual heavy elements.
To quantitatively predict the difference between the elemental abundances
of clouds enriched by uniform and stratified ejecta, it is essential to consider radial
distribution of each heavy element.
Here we run simulations with a wide range of explosion energies $\Esn = 0.7$--$10$ B with
ejecta models taken from SN nucleosynthesis calculations that give an accurate
radial distribution of heavy elements for each explosion energy.

Due to computational limits, we start the simulations at a time when the ejecta has 
expanded up to a radius $\Rs \simeq 3$ pc.
The corresponding elapsed time from the SN explosion is 
$9.0\E{3} \ {\rm yr} \left( \Rs / 3 \ {\rm pc} \right)^{5/2} 
\left( \Esn / 1 \ {\rm B} \right)^{-1/2}$ \citep{McKee77}.
Unfortunately, the mixing efficiency of ejecta in the early phase ($<1000$ yr) is unknown.
During the shock propagation in the stellar mantle,
the layers in ejecta with different mean molecular weights can mix
with each other between their boundaries because of Rayleigh-Taylor (RT) instabilities
\citep{Joggerst10}.
Observations of the young ($\simeq 1000$ yr) SN remnant (SNR) Cas A showed a knotty elemental 
distribution \citep{Douvion01},
and X-ray/$\gamma$-ray observations also showed that the ejecta of SN 1987A is partially mixed \citep{Dotani87}.
These observations suggest that the mixing state is somewhere between fully stratified and
mixed.
We therefore run simulations in two extreme cases, stratified and uniform ejecta.
Our results give the upper and lower limits of its effect on the elemental abundances
of the clouds that host second-generation stars.

The structure of this paper is as follows.
In Section \ref{sec:method}, we describe our numerical methods.
We present our results in Section \ref{sec:results} and compare them 
with the elemental abundance in EMP stars in Section \ref{sec:discussion}.
Finally, we conclude in Section \ref{sec:conclusion}.
Throughout this paper, the simulations are performed in comoving coordinates, but
we quote proper values unless otherwise specified.

%%%%%%%%%%%%%%%%%%%%%%%%%%%%%%%%%%%%%%%%%%%%%
% Numerical Simulations %%%%%%%%%%%%%%%%%%%%%
%%%%%%%%%%%%%%%%%%%%%%%%%%%%%%%%%%%%%%%%%%%%%

\begin{figure*}
\includegraphics[width=18.0cm]{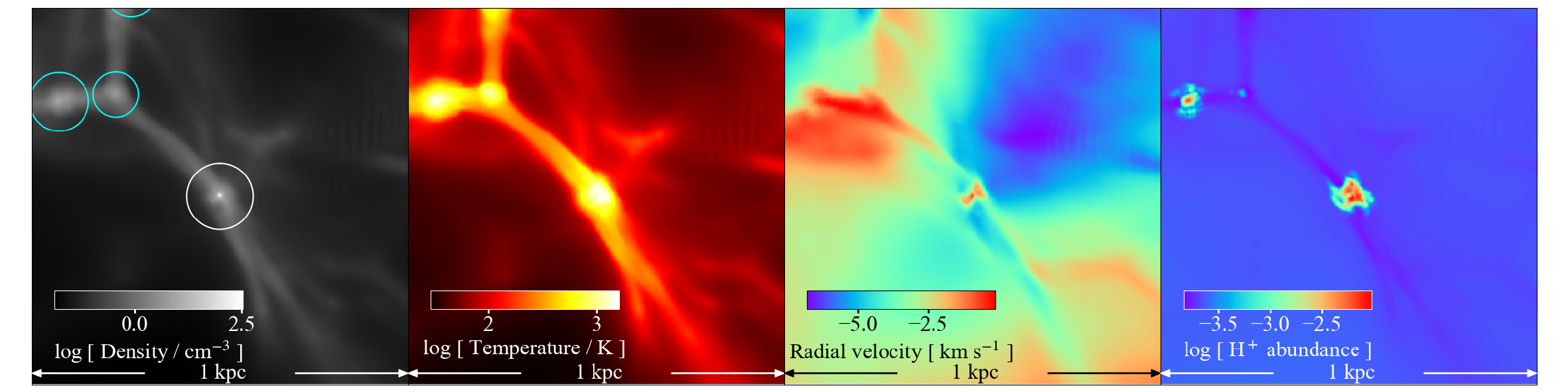}
\caption{
Density-weighted projection of density, temperature, and radial velocity relative to the central Pop III star,
and H {\sc ii} abundance relative to hydrogen nuclei
at the end of the lifetime of the progenitor star with
mass $\MPopIII = 25 \ \Msun$, just before its supernova explosion.
The white and cyan circles depict the virial radii of the Pop III hosting halo which is eventually internally enriched 
and halos which will merge and be externally enriched for an explosion energy $\Esn = 10$ B.
}
\label{fig:snapshots_ini}
\end{figure*}

\section{Numerical models}
\label{sec:method}

In this work, we use basically the same numerical method as in \citetalias{Chiaki18}.
We briefly summarize our basic setup in Sections 2.1--2.2 
and detail the updated methods in Section 2.3.

\subsection{Simulation setup}

We utilize the SPH/$N$-body hydrodynamics code {\sc gadget}-2 \citep{Springel05}, while considering 
non-equilibrium chemistry and cooling.
The hydrodynamics of the gas component is solved with the standard SPH scheme with a fixed number of
neighboring particles $\Nngb = 64\pm 8$.
To alleviate spurious surface tension on contact discontinuities
\citep{Saitoh13, Hopkins13, Wadsley17},
we adopt the shared timestep strategy, where the physical variables of all SPH particles 
are updated with a global timestep \citep{Saitoh09}.
The timestep is calculated as 
$\underset{i}{\min} \{\Delta t_{{\rm acc},i}, \Delta t_{{\rm cou},i} \}$, where
$\Delta t_{{\rm acc},i}$ and $\Delta t_{{\rm cou},i}$ are the acceleration and Courant timescales
of a gas particle $i$, respectively.

We solve the chemical networks of 53 reactions for 15 species,
e$^-$, H$^+$, H, H$^-$, H$_2^+$, H$_2$, D$^+$, D, D$^-$, HD$^+$, HD, He$^{2+}$, He$^+$, He, and HeH$^+$.
Our chemical model includes collisional ionization/recombination of H/He and
the H$^-$-process, H$_2^+$-process, and three-body reactions for the formation of molecular hydrogen.
We then calculate the cooling rates of inverse Compton, bremsstrahlung,
ionization/recombination and collisional excitation of H/He,
and ro-vibrational transition of H$_2$/HD molecules.
We here ignore C and O fine-structure cooling, which is dominated by
HD cooling for metallicities $\abH{C} \lesssim -3$ \citep{Omukai05, Jappsen07}.

We follow the dispersion of metals with Lagrangian tracer particles.
The velocity ${\vb v}_j$ of a tracer particle $j$ are interpolated from 
the velocity ${\vb v}_i$ of neighboring SPH particles as 
\begin{equation}
{\vb v}_j = \sum _i \frac{m_i }{ \rho _i } {\vb v}_i W(r_{ij}, \tilde h_j),
\label{eq:velocity}
\end{equation}
where $m_i$ and $\rho _i$ are the mass and density of a gas particle $i$, respectively, and
$W(r_{ij}, \tilde h_j)$ is the cubic-spline kernel function of
the distance $r_{ij}$ between particles $i$ and $j$ \citep{Springel05}.
The smoothing length $\tilde h_j$ is estimated so that a sphere with the radius $\tilde h_j$ contains
four neighboring SPH particles.
The number of neighboring particles for the velocity evaluation is smaller 
than the density evaluation to capture steep velocity gradients around SN shocks.

\subsection{Initial conditions}

We create initial conditions for a cosmological zoom-in simulation
with the {\sc music} code \citep{Hahn11}.
We initialize the simulation at $z_{\rm ini} = 99$ with cosmological parameters taken from
\citet{Planck2015}.
First, we run a dark matter simulation with $512^3$ particles in a $10 h^{-1}$ Mpc (comoving)
periodic box.
We refine dark matter particles around MHs identified with a friends-of friends (FOF) algorithm.
Then, we restart the simulation from $z_{\rm ini}$ including baryons.
The minimum mass of dark matter and baryon particles are $m_{\rm p, gas} = 4.45 \ \Msun$ and
$m_{\rm p, dm} = 16.5 \ \Msun$, respectively.
A MH forms at redshift $z_{\rm form} = 28.5$ with a
virial radius $\Rvir = 70.1$ pc and mass $\Mvir = 3.33\E{5} \ \Msun$.
We cut out a spherical region with a radius 2.5 kpc centered on the center-of-mass of the MH,
which contains 9,994,502 dark matter particles and 9,972,070 gas particles.

When the maximum density reaches $\nHcen = 10^3 \ \percc$,
we put a star particle with a mass $\MPopIII = 25 \ \Msun$ 
at the center-of-mass of a region with densities $> \nHcen/3$.
We solve radiation transport from the star with
a scheme of \citet{Susa06}.
The emission rate of hydrogen ionization photons is set to
$Q({\rm H}) = 7.58\E{48} \ {\rm s}^{-1}$ \citep{Schaerer02}.
Fig. \ref{fig:snapshots_ini} shows the density, temperature, and radial velocity of the MH,
and H {\sc ii} abundance relative to hydrogen nuclei
at the end of the lifetime of the star $\tlife = 6.46$ Myr, corresponding to a redshift $z = 27.3$.
The virial mass and radius of the halo grow to $\Mvir = 5.11\E{5} \ \Msun$ and $\Rvir = 87.6$ pc, 
respectively, through smooth mass accretion.
The gas is partially ionized within $\sim 1$ pc from the star, where
density, temperature, and H$^+$ abundance are
$\nH \sim 300 \ \percc$, $T \sim 5000$ K, and $y({\rm H^+}) \sim 0.1$, respectively.
The radius of the H {\sc ii} region is smaller than the halo virial radius because 
the expansion of ionizing front is halted by the gas accretion onto the MH 
\citep[see also][]{Kitayama04}.\footnote{The simulation is the same as the {\tt MH1-C25} 
run in \citetalias{Chiaki18} until the SN explosion occurs.}

Note that the radius of the 
ionized region in this simulation is smaller than the Str\"{o}mgren radius 
$R_{\rm St} = 36 \left( Q({\rm H}) / 8\E{48} \ {\rm s}^{-1} \right) \left( n_{\rm e} / 1 \ \percc \right) ^{-2/3} \ {\rm pc}$
because of the resolution limit as we discuss in Section 4.2.1.
Also, we here consider the formation of a single Pop III star in the MH
although the fragmentation of accretion disks can lead to
multiple Pop III star formation 
\citep[e.g.,][]{Turk09, Clark11, Greif12}.
In addition, we prevent star formation in the other MHs by
switching off gas cooling in dense regions with $> 10^3 \ \percc$ 
at distances $> 100$ pc from the central MH.
Lastly, we do not consider a relative velocity offset between dark matter
and baryons (``streaming velocity'').
This would play an important role in the structure formation and star formation
\citep{Chiou18, Chiou19, Druschke19}.
The validity of these assumptions is discussed in Section \ref{sec:multiPopIII}.

\begin{figure*}
\includegraphics[width=\textwidth]{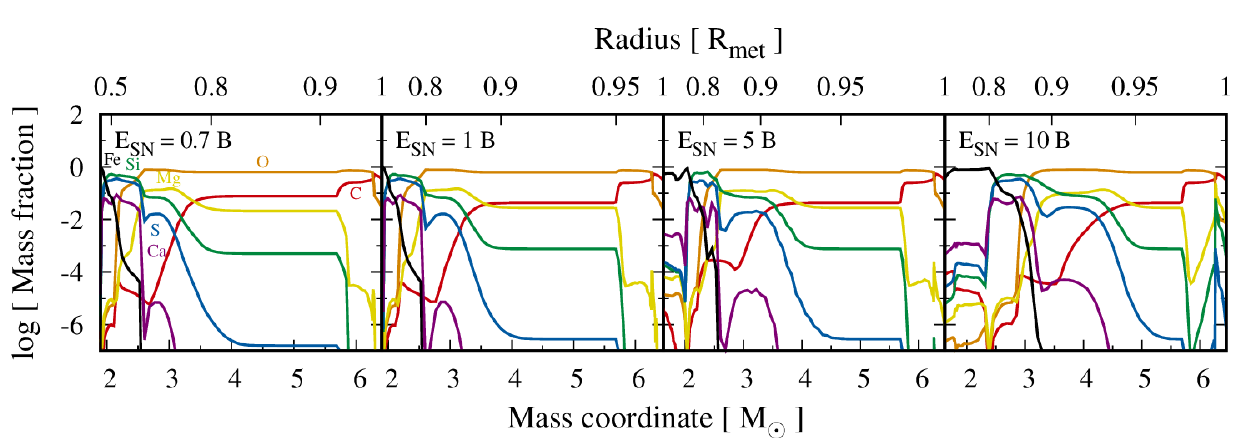}
\caption{Mass fraction of major elements C (red), O (orange), Mg (yellow), Si (green),
S (blue), Ca (purple), and Fe (black) as a function of enclosed mass
and radius of ejecta for our progenitor models
with a mass $\MPopIII = 25 \ \Msun$ and explosion energy $\Esn = 0.7$, 1, 5, and 10 B
($1 \ {\rm B} = 10^{51} \ \erg$).
We plot the results of nucleosynthesis calculations of \citet{Tominaga14}
only in the CO core (see text).
The top $x$-axis shows the spatial scale of the ejecta normalized
by the radius $R_{\rm met}$ of the CO core.}
\label{fig:mx_ini}
\end{figure*}

\begin{table*}
\begin{minipage}{13.5cm}
\caption{SN models}
\label{tab:SN}
\begin{tabular}{cccccccccc}
\hline
$^{1}$Run  & $^{2}\Esn$      & $^{3}M_{\rm rem}$ & $^{4}M_{\rm met}$ & $^{5}R_{\rm met}$ & $^{6}R_{\rm eje}$ & $^{7}M_{\rm C}$ & $^{8}M_{\rm O}$  & $^{9}M_{\rm Fe}$ & $^{10}$[C/Fe]$_0$    \\
           & $[{\rm B}]$     &  $[\Msun]$        &  $[\Msun]$        &   [$10^{-6}$ pc]  &   [$10^{-6}$ pc]  & $[\Msun]$       & $[\Msun]$        & $[\Msun]$        &                      \\
\hline \hline                                                                                                                                                                            
{\tt E0.7} & $ 0.7$          & $ 1.90 $          & $4.51$            & $1.14$           & $19.4$             & $0.384$         & $2.61$           & $0.0689$         & $ 0.48$              \\
{\tt E1}   & $ 1$            & $ 1.90 $          & $4.55$            & $1.89$           & $51.6$             & $0.331$         & $2.58$           & $0.0885$         & $ 0.31$              \\
{\tt E5}   & $ 5$            & $ 1.62 $          & $4.83$            & $5.79$           & $47.6$             & $0.329$         & $2.61$           & $0.271 $         & $-0.18$              \\
{\tt E10}  & $10$            & $ 1.62 $          & $4.83$            & $8.42$           & $66.7$             & $0.285$         & $2.23$           & $0.699 $         & $-0.65$              \\
\hline
\end{tabular}
\medskip \\
Note --- (1) ID of runs. 
(2) Explosion energy.
(3--4) Mass of a compact remnant and CO core.
(5--6) Radius of CO core and ejecta.
(7--9) Mass of synthesized C, O, and Fe.
(10) Average abundance ratio $\abFe{C}$ in the CO core.
\end{minipage}
\end{table*}

\begin{figure*}
\includegraphics[width=14.0cm]{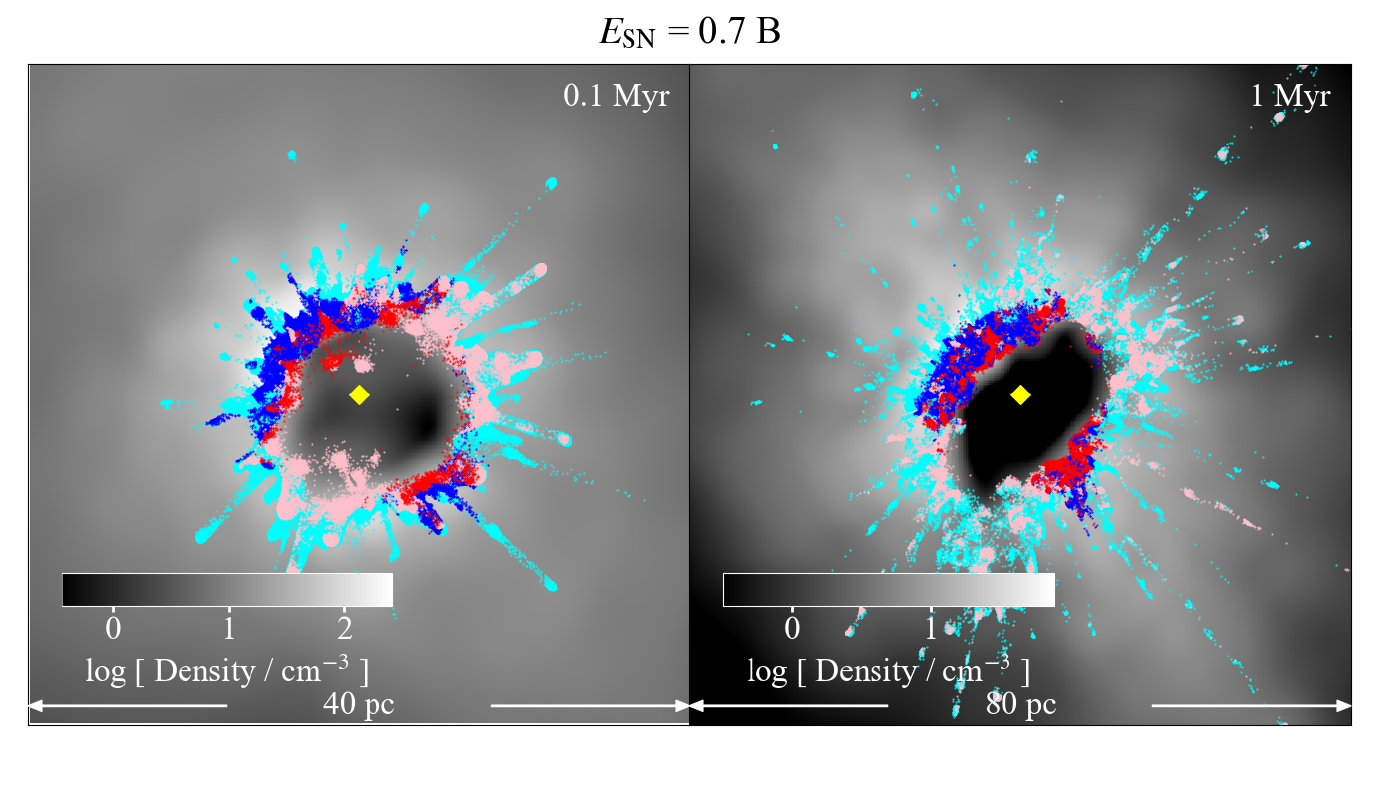}
\caption{
Slices of density 0.1 Myr (left panel) and 1 Myr (right panel) after the supernova 
for an explosion energy $\Esn = 0.7$ B.
The panels are centered on the Pop III remnant (yellow diamond).
We over-plot the location of lighter and heavier element dominant metal particles with $\abX{C} > \abX{Fe}$ (blue dots)
and with $\abX{Fe} > \abX{C}$ (red dots), respectively, in a slab with the depth of 0.1 times the window size, 
where $\abX{M}$ is the mass fraction of metal element M.
The dots with darker colors represent the particles which eventually fall back into 
the recollapsing cloud center ($r < \Rjeans$).
}
\label{fig:snapshots_lowE}
\end{figure*}

\begin{figure*}
\includegraphics[width=14.0cm]{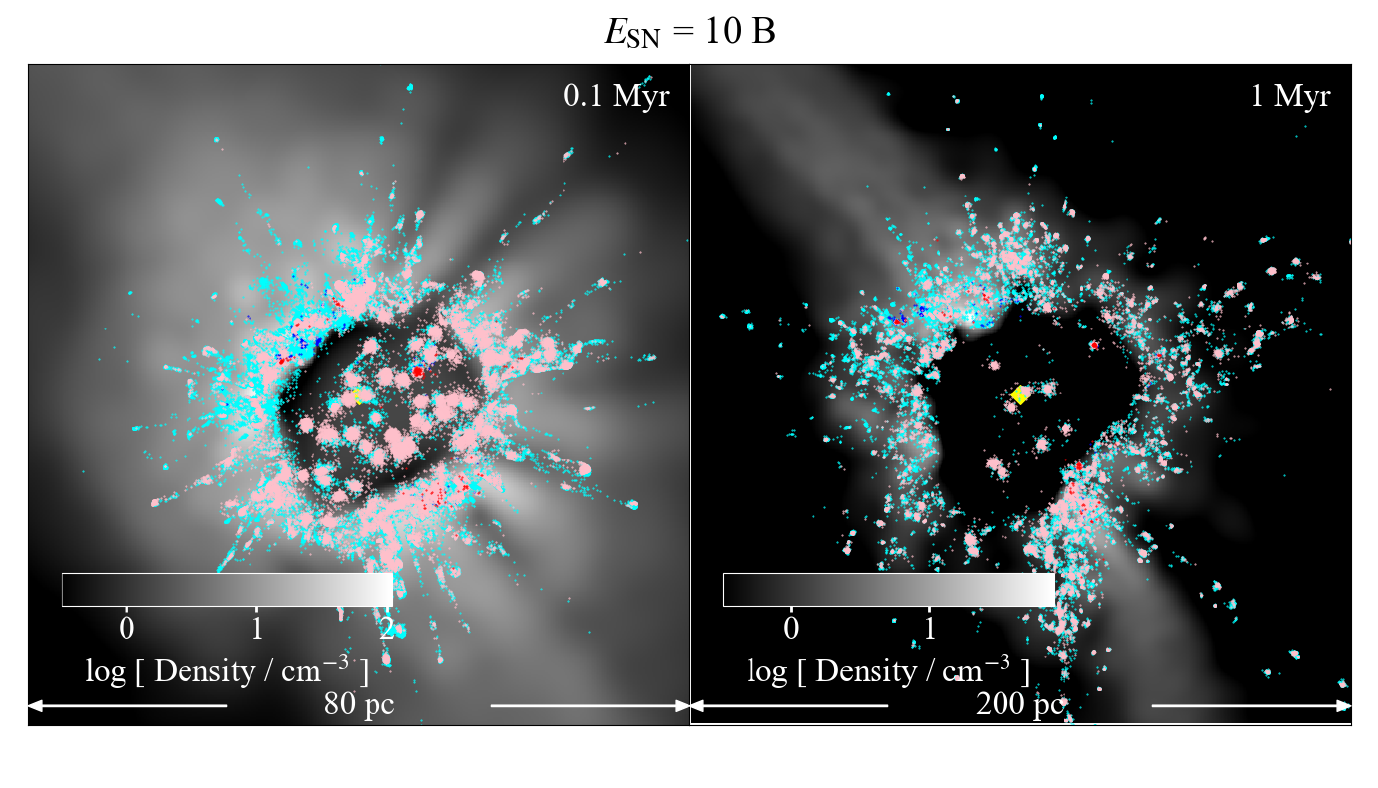}
\caption{
Same as Fig. \ref{fig:snapshots_lowE} but for an explosion energy $\Esn = 10$ B.
}
\label{fig:snapshots_highE}
\end{figure*}

\subsection{Supernova models}
\label{sec:SNmodels}

From the snapshot at $\tlife$ (Fig. \ref{fig:snapshots_ini}),
we run four simulations with explosion energies $\Esn = 0.7$,
$1$, $5$, and $10$ B ($1 \ {\rm B} = 10^{51} \ \erg$), 
hereafter called as {\tt E0.7}, {\tt E1}, {\tt E5}, and {\tt E10}, respectively.
We replace the star particle with a Pop III
remnant particle with a mass $M_{\rm rem}$ (Table \ref{tab:SN})
and uniformly inject the explosion energy $\Esn$ to central 200 SPH particles 
in a sphere with a radius $R_{\rm eje, sim} = 2.77$ pc.
We deposit all of the explosion energy as thermal energy.

In this region, we insert $10^6$ metal particles,
and the elemental mass fractions of each particle
are remapped from one-dimensional stellar evolution/nucleosynthesis models
of \citet{Tominaga14}.
In their models, 
heavier elements than He are contained in only a small part of the ejecta.
The metal-rich region with $\abX{H} + \abX{He} < 0.95$ (hereafter called ``CO core'')
lies only within the radius $R_{\rm met}$, at least $\sim 4$\% of entire ejecta radius $R_{\rm eje}$ (Table \ref{tab:SN}),
where $\abX{M}$ is the mass fraction of an element M.
If we include the all part of ejecta,
the velocity of metal particles in the CO core would be interpolated from
only $200 \times 0.04^3 = 0.01$ SPH particles (Table \ref{tab:SN}).
Therefore, we consider only the CO core in our simulations.
Fig. \ref{fig:mx_ini} shows the mass fractions of major elements 
in the CO core calculated by \citet{Tominaga14}.

We note that 2.34\%, 13.1\%, 13.1\%, and 14.8\% of C is produced in the outer region
containing mainly primordial elements (called ``hydrogen envelope'') for 
{\tt E0.7}, {\tt E1}, {\tt E5}, and {\tt E10}, respectively.
This means 0.01--0.07 dex loss of carbon mass.
We also note that only $1.18\E{-5}$, $2.09\E{-4}$, $3.65\E{-4}$, and $8.66\E{-4}$ of N is produced in the CO core.
Thus, even without the hydrogen envelope, the radial velocity of the CO core is not overestimated 
because of the deceleration from the swept up materials.

Table \ref{tab:SN} shows the CO core mass $M_{\rm met}$ and
C, O, and Fe mass ($M_{\rm C}$, $M_{\rm O}$, and $M_{\rm Fe}$, respectively).
The dominant element is O for all $\Esn$.
The mass of lighter elements decreases while the mass of heavier elements 
increases with increasing $\Esn$.
The elemental abundance ratio of the uniform ejecta, which is often used when fitting
EMP stars, are calculated from the total mass of each element produced in the ejecta.
We hereafter attach the index `0' to depict the average elemental abundance ratios.
In our normal CCSN model, the average carbon-to-iron abundance ratio $\abFe{C}_0$ increases from $-0.65$
for {\tt E10} to $0.48$ for {\tt E0.7},
below the definition of CEMP stars ($\abFe{C} = 0.7$).
Note that 80\% of EMP stars are C-enhanced stars with metallicities $\abH{Fe} < -4.0$ \citep{Yoon18, Norris19}.
We discuss the metal enrichment scenario from a different SN model which can explain C-enhanced 
star formation in Section \ref{sec:CEMP}.

We terminate the simulations at the time $\trecol$ when the maximum density of 
an enriched gas cloud reaches $\nHcen = 10^3 \ \percc$.
We compare the elemental abundance within half of a Jeans length 
\begin{equation}
\Rjeans = \frac{1}{2} \left( \frac{\pi \cs ^2}{G \rho} \right) ^{1/2}
\end{equation}
with the abundances from uniformly mixed ejecta.

\begin{figure*}
 \includegraphics[width=18.0cm]{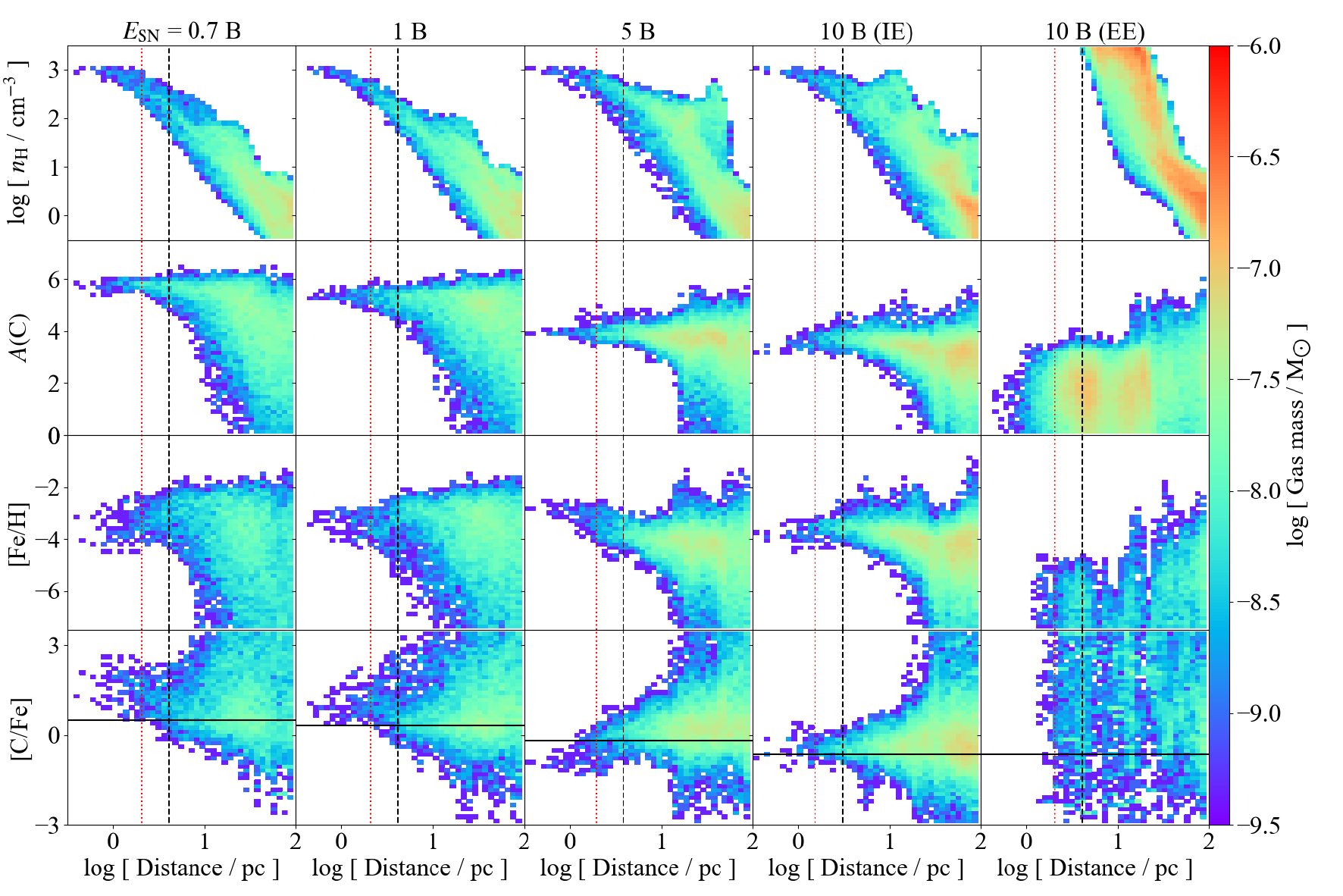}
\caption{
Two-dimensional histogram of density, $\abA{C}$, $\abH{Fe}$, and $\abFe{C}$ 
as a function of radius from the density maximum at the time when the density of
the enriched clouds reaches $10^{3} \ \percc$.
The horizontal black solid lines show $\abFe{C}$ with uniform ejecta.
The vertical black dashed and red dotted lines show half of the Jeans length of the recollapsing clouds
and resolution limit, respectively (see text).
}
\label{fig:rn_fin}
\end{figure*}

\begin{figure}
 \includegraphics[width=8.5cm]{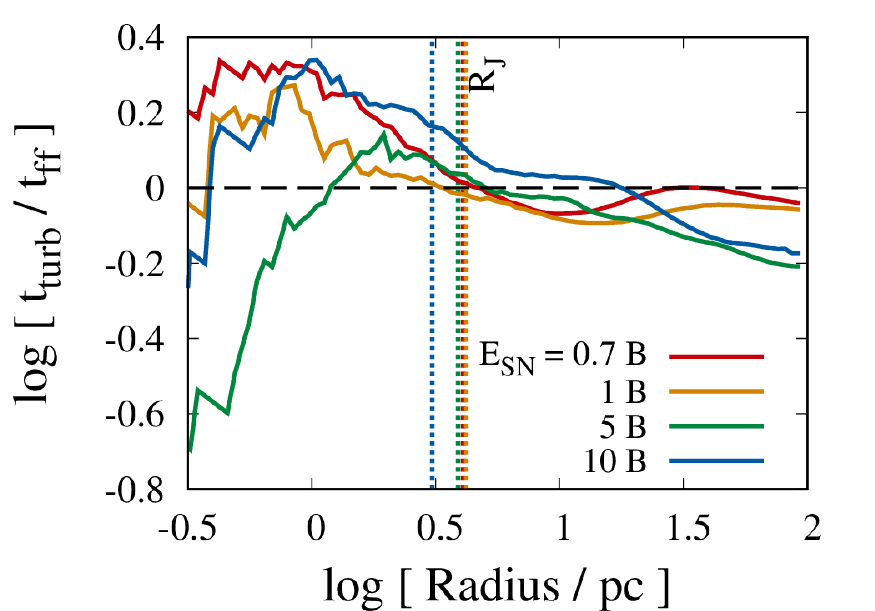}
\caption{
Ratio of the timescale of turbulent mixing $\tturb$ to free-fall time $\tff$ 
as a function of radius at the time $\trecol$ when the density of recollapsing clouds reaches $10^{3} \ \percc$ (solid curves). 
The dotted lines show half of the Jeans length $\Rjeans$.
The colors depict explosion energies with 
$\Esn = 0.7$ B (red), $1$ B (orange), $5$ B (green), and $10$ B (blue).
}
\label{fig:rtt_fin}
\end{figure}

\begin{figure*}
\includegraphics[width=\textwidth]{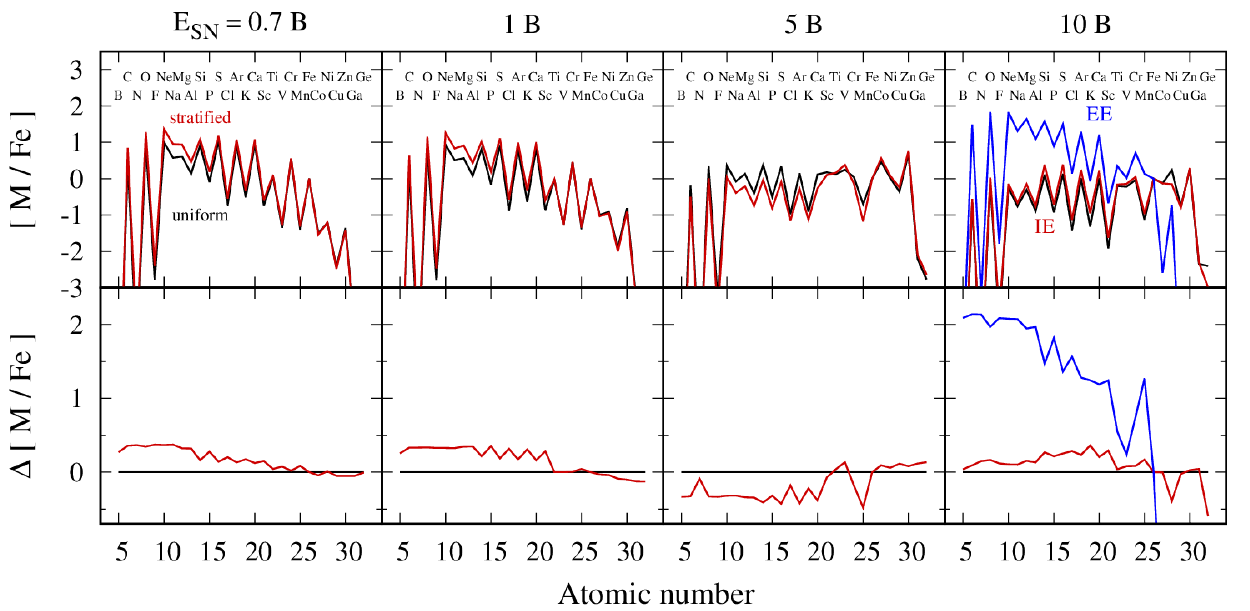}
\caption{
Abundance ratios of element M from B to Ge relative to iron in the internally (red) and externally (blue) 
enriched clouds from the stratified ejecta for explosion energies $\Esn = 0.7$ B, 1 B, 5 B, and 10 B 
from left to right.
The black curves show the elemental abundance of the uniform ejecta.
}
\label{fig:M_Fe}
\end{figure*}

%%%%%%%%%%%%%%%%%%%%%%%%%%%%%%%%%%%%%%%%%%%%%
% RESULTS %%%%%%%%%%%%%%%%%%%%%%%%%%%%%%%%%%%%%%
%%%%%%%%%%%%%%%%%%%%%%%%%%%%%%%%%%%%%%%%%%%%%

\section{Results}
\label{sec:results}

\subsection{Overview}

Figs. \ref{fig:snapshots_lowE} and \ref{fig:snapshots_highE} show the gas density
at 0.1 and 1 Myr after the SN explosion for {\tt E0.7} and {\tt E10}.
The blue and red dots depict the distribution of metal particles
with $\abX{C} > \abX{Fe}$ and $\abX{C} < \abX{Fe}$, respectively.
In all progenitor models, a part of ejecta falls back mainly along the
cosmological filaments after the SN shells lose its thermal energy through radiative cooling
in the dense contact surfaces with the filaments (Fig. \ref{fig:snapshots_ini}).
The gas falling back into the MH begins to collapse again, i.e., 
internal enrichment (IE) occurs at $\trecol = 5.69$--37.2 Myr for different explosion energies (Table \ref{tab:metallicity}).
The darker colors in Figs. \ref{fig:snapshots_lowE} and \ref{fig:snapshots_highE}
depict the particles which eventually fall back into the MH.
These particles are confined in the contact surfaces.

Fig. \ref{fig:rn_fin} shows the radial distribution of density, $\abA{C}$, $\abH{Fe}$, and
$\abFe{C}$ of the gas particles as a function of distance from the density maximum of the enriched clouds.
We also show half of the Jeans length $\Rjeans$ (black dashed line) and the resolution limit defined
as twice the smoothing length (red dotted line).
The carbon and iron abundances show large deviations at distances $> 10$ pc because of the non-uniform 
metal mixing in the voids.
Toward the higher density region at distances $< 3$ pc, the metal abundances converge.
In Fig. \ref{fig:rtt_fin} we compare the timescale for metal mixing due to turbulence, 
$\tturb = R / \sigma _v$, with the free-fall timescale, $\tff = (3\pi/32G\bar \rho)^{1/2}$,
where $\sigma _v$ and $\bar \rho$ are the velocity dispersion and mean density
within a radius $R$, respectively.
Within $R \lesssim 3$ pc the mixing timescale becomes longer than the collapsing timescale,
which indicates that further metal mixing does not occur
as pointed out by earlier studies \citep{Smith15, Ritter15, Chiaki19}.

We measure the average abundance of each element within $\Rjeans \sim 3$ pc,
regarding it as the characteristic abundance of the clouds.
The red and black curves in Fig. \ref{fig:M_Fe} show the elemental abundance
ratios $\abFe{M}$ 
for the stratified and uniform ejecta, respectively.
The difference $\Delta \abFe{M}$ of these abundances is within $\pm 0.4$ dex 
for all elements for all progenitor models (bottom panels of Fig. \ref{fig:M_Fe}).
We can conclude that the stratified structures of the ejecta hardly affect on the elemental abundance ratio
of the internally enriched clouds, which we will further analyze in Section 3.2.1 (i).

\begin{figure}
\includegraphics[width=8.5cm]{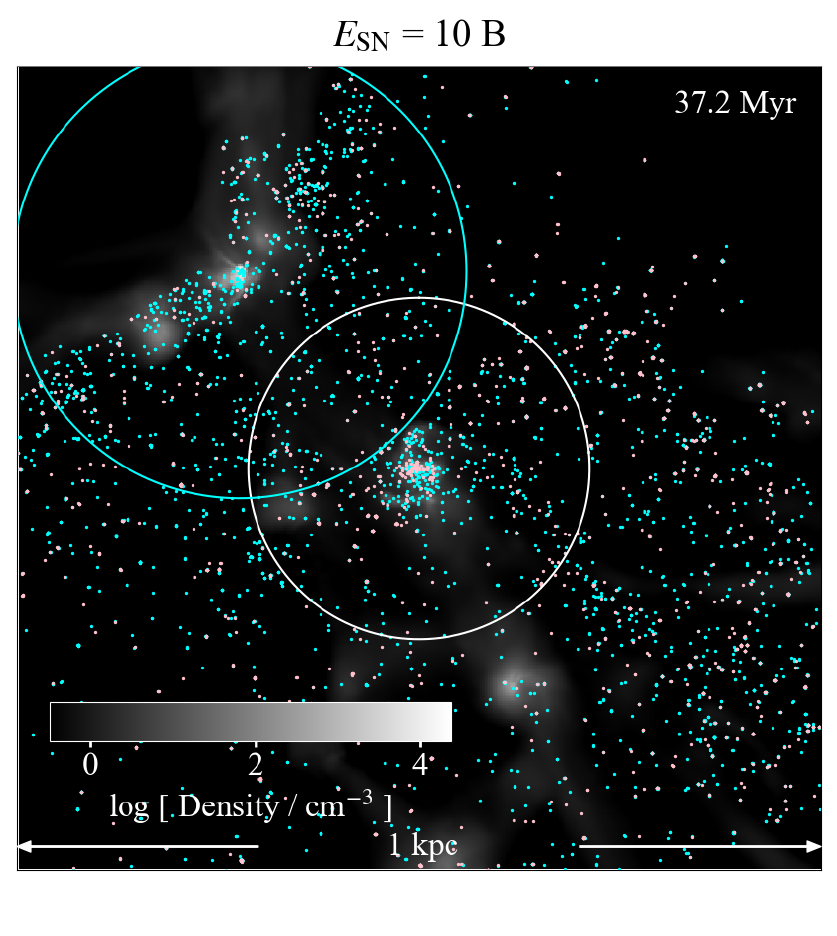}
\caption{
Density-weighted projection of density for an explosion energy $\Esn = 10$ B and location of
C-dominated (cyan) and Fe-dominated (pink) metal particles.
We plot every 100 particles.
The white and cyan circles denote virial radii of the Pop III hosting halo and externally enriched halo.
}
\label{fig:snapshots_E10_EE}
\end{figure}

\begin{figure*}
\includegraphics[width=\textwidth]{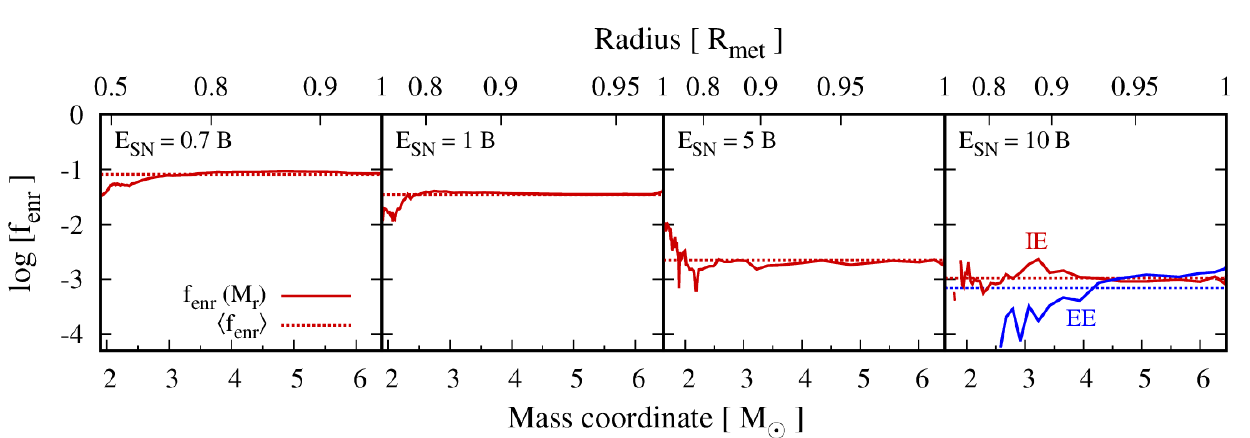}
\caption{Radial profile of mass fraction $\ffb$ of metals incorporated into the subsequent star-forming 
clouds at the time when the gas density reaches $10^{3} \ \percc$. 
The positions of metals are traced back to the initial mass coordinate and radius of ejecta
(the same axes as Fig. \ref{fig:mx_ini}).
The red and blue curves depict the results for internal and external enrichment, respectively.
We show $\ffb (\Mr)$ as a function of mass coordinate $\Mr$ (solid curves) and 
their mass-weighted average $\langle \ffb \rangle$ among radial bins (dotted curves).}
\label{fig:mf_ini}
\end{figure*}

For {\tt E10}, metals reach the neighboring halo, and external enrichment (EE) occurs (Fig. \ref{fig:snapshots_E10_EE})
as well as IE.
The blue curve of Fig. \ref{fig:mf_ini} shows that only the outer layers of ejecta with mass coordinates $M_r > 2.5 \ \Msun$
reach the neighboring halo.
Since these layers mainly contain lighter elements (bottom panel of Fig. \ref{fig:M_Fe}),
the relative abundances of lighter elements ($< {\rm Ca}$) are enhanced more than $\Delta \abFe{M} > 1$.
This indicates that stars with the C-enhanced abundance ($\abFe{C} = 1.49$) can form in externally enriched
clouds from normal CCSNe.

\begin{table*}
\begin{minipage}{18.0cm}
\caption{Metallicity in enriched clouds}
\label{tab:metallicity}
\begin{tabular}{cccccccccccc}
\hline
$^{1}\Esn$      & $^{2}\trecol$ & $^{3}\zrecol$ & $^{4}$enrich.  &  $^{5}\Mvir$    & $^{6}\Rvir$  &  $^{7}\Mcl$     & $^{8}R_{\rm J}$ & $^{9}\abA{C}$ & $^{10}\abH{Fe}$ & $^{11}$[C/Fe]  & $^{12}\Delta \abFe{C}$\\
$[{\rm B}]$     & [Myr]         &               &                &  [$\Msun$]      &  [pc]        &  [$\Msun$]      & [pc]            &               &                 &                &                       \\
\hline \hline                                                                                                                                                                                                    
$ 0.7$          &  5.69         &  26.4         & IE             & $7.25\E{5}$     &  102         & $1.76\E{3}$     & $4.08$          & $6.32$        & $-2.95$         & $ 0.84$        & $ 0.36$               \\
$ 1.0$          &  7.30         &  26.2         & IE             & $7.82\E{5}$     &  105         & $1.81\E{3}$     & $4.19$          & $5.86$        & $-3.21$         & $ 0.64$        & $ 0.33$               \\
$ 5.0$          & 24.7          &  23.9         & IE             & $2.02\E{6}$     &  158         & $2.01\E{3}$     & $3.89$          & $4.56$        & $-3.36$         & $-0.50$        & $-0.33$               \\
$10.0$          & 37.2          &  22.5         & IE             & $4.13\E{6}$     &  212         & $1.39\E{3}$     & $3.06$          & $4.32$        & $-3.55$         & $-0.56$        & $ 0.09$               \\
                &               &               & EE             & $9.73\E{6}$     &  282         & $1.40\E{5}$     & $4.09$          & $2.48$        & $-7.44$         & $ 1.49$        & $ 2.14$               \\
\hline
\end{tabular}
\medskip \\
Note --- 
(1) Explosion energy.
(2) Time of recollapse from SN explosion.
(3) Redshift of recollapse.
(4) Enrichment mode.
(5--6) Virial mass and radius of the MH at the time of recollapse.
(7--8) Mass and half of the Jeans length of the recollapsing cloud.
(9--11) Average carbon and iron abundances and relative abundance ratio $\abFe{C}$ of the recollapsing cloud.
(12) Difference of $\abFe{C}$ between the stratified and uniform ejecta.
\end{minipage}
\end{table*}

\subsection{Evolution of SN shells}

\subsubsection{Internal enrichment}
\label{sec:IE}

\noindent{\bf (i) General behavior}

In the case of IE, there is only a slight difference ($\pm 0.4$ dex) in the elemental
abundance ratio between the clouds enriched by the stratified and uniform ejecta.
Fig. \ref{fig:mf_ini} shows the mass fraction $\ffb (\Mr)$ of metals falling 
back into the MH as a function of initial mass mass coordinate $\Mr$ of the ejecta.
For all $\Esn$, $\ffb (\Mr)$ is almost constant against $\Mr$.
Therefore, the ratio between lighter and heavier elements does not significantly deviate from that for
the uniform ejecta.

In this section, we interpret this property by reviewing the evolution of SN shells.
First, in the Sedov-Taylor (ST) phase, the shell evolves adiabatically because the timescale of radiative cooling 
is longer than the elapsed time.
Then the shell enters the pressure-driven snowplough (PDS) phase, in which the dense cooling shell is
pushed by the pressure in a hot inner cavity.
Because of the inertial force from the decelerating shell to the ISM, RT instabilities
develops and induces the mixing of materials between the shell and ISM.
After the internal cavity cools down through radiative cooling, 
the shell continues to expand with its momentum being conserved
in the phase called momentum-conserving snowplough (MCS) phase \citep{Ostriker88}.
The shell eventually either dissolves into the ISM if the explosion energy is enough high
or otherwise falls back into the MH \citep{Ritter12, Sluder16, Smith15, Chiaki19}

Figs. \ref{fig:snapshots_lowE} and \ref{fig:snapshots_highE} show that
lighter and heavier elements are both concentrated
and partly mixed in the dense cooling shells which have developed
in the PDS phase.
As depicted by the dots with darker colors,
both lighter and heavier elements that eventually fall back into the MH
are confined in the same region where the SN shell interacts with
the dense cosmological filaments.
Using the radius of the filaments $\Rf$ and the radius of the SN shell $\Rs$, 
we can estimate the fall back fraction as the ratio of the
solid angles $\ffb = \pi \Rf ^2 / 4 \pi \Rs ^2$ \citep{Ritter15}.
From this, the abundances of C and Fe which fall back can be estimated as
\begin{eqnarray}
\abA{C}  &=& 12 + \log \left( \frac{\ffb \abM{C}  / \muX{C} }{\XH \Mcl} \right), \label{eq:AC} \\
\abH{Fe} &=& 12 + \log \left( \frac{\ffb \abM{Fe} / \muX{Fe}}{\XH \Mcl} \right) - \abAsun{Fe} \label{eq:FeH},
\end{eqnarray}
where $\muX{M}$ is the molecular weight of an element M ($\muX{C} = 12$, $\muX{Fe} = 56$)
and $\abAsun{Fe}=7.50$ is the solar abundance of Fe \citep{Asplund09}.
The mass $\Mcl$ is defined as the Jeans mass of the recollapsing clouds.
From these equations, the abundance ratio
\begin{equation}
\abFe{C} = \left(\abA{C} - \abAsun{C}\right) - \abH{Fe}
\end{equation}
is the same as that for the uniform ejecta 
\begin{equation}
\abFe{C} _0 = \log \left( \frac{\abM{C}  / \muX{C} }{\abM{Fe} / \muX{Fe}} \right) 
\end{equation}
because the factor $\log \ffb$ is canceled out.
Since the factor depends only on 
the geometry of the circumstellar medium,
the fall back fraction of each elements is almost the same.

For {\tt E0.7} and {\tt E5}, the shell radius in the direction of filaments is
$\Rs = 20$ pc and 60 pc, and
the radius of the dense part of the filaments is 
$\Rf = 10$ pc and $5$ pc, respectively.
Then, the enrichment fraction is
$\ffb = 6.25\E{-2}$ and $1.74\E{-3}$.
This is consistent with the average fall back fraction
estimated from our simulations ($8.12\E{-2}$ and $2.23\E{-3}$; dotted lines in Fig. \ref{fig:mf_ini}).
With the typical mass of recollapsing clouds $\Mcl = 2000 \ \Msun$ (Table \ref{tab:metallicity}), 
the abundances can be estimated to be
($\abA{C}$, $\abH{Fe}$) = ($6.12$, $-2.80$) and
($4.50$, $-3.76$) for {\tt E0.7} and {\tt E5}, respectively.
These estimates are also consistent with the simulation results.

\vspace{0.5cm}
\noindent{\bf (ii) Dependency of elemental abundance on $\Esn$}

The difference of the abundance ratio of lighter to heavier elements
between the stratified and uniform ejecta depends slightly on $\Esn$.
The bottom panels of Fig. \ref{fig:M_Fe} show that the ratio is higher than that from the uniform 
ejecta for lower explosion energies $\Esn \leq 1$ B ($0.33 < \Delta \abFe{C} < 0.36$) while similar
or smaller for higher $\Esn \geq 5$ B ($-0.33 < \Delta \abFe{C} < 0.09$).
The shell evolution and elemental abundances in the enriched clouds mostly differ between the two cases
with explosion energies $\Esn \leq 1$ B and $\Esn \geq 5$ B, hereafter called
low and high $\Esn$, respectively.
We compare these two cases in this section.

For low $\Esn$, the MCS phase begins at $\sim 0.1$ Myr after the SN explosion.
In the MCS phase, the contact surface of the shell in the direction of filaments is 
decelerated from the pressure of infalling gas along the filaments with a radial velocity 
$\sim 6 \ {\rm km \ s^{-1}}$ (Fig. \ref{fig:snapshots_ini}).
In comparison, the hot inner part of the shell containing the heavier elements 
circumvents the dense contact surface and continues to expand in the direction of voids.
Therefore, the abundance of Fe falling back into the cloud is slightly lower 
than the simple estimation of Eq. (\ref{eq:FeH}).
As a result, the abundance ratio of lighter to heavier elements from the stratified ejecta becomes
slightly larger than that from the uniform ejecta ($0.33 < \Delta \abFe{C} < 0.36$).
This is the consistent result with the high-resolution simulation of \citet{Ritter15} and \citet{Sluder16}.

For high $\Esn$, the shell reaches a larger distance ($\sim 60$ pc) and
propagates into the less dense region ($\lesssim 1 \ \percc$) than for low $\Esn$ (Fig. \ref{fig:snapshots_highE}).
The pressure from the filaments on the contact surface is smaller 
and the outer part of the shell is decelerated less efficiently.
The inner part of the shell moves radially without circumventing the contact surface.
Therefore, the deviation of the abundance ratio between lighter and heavier elements 
falling back into the MH is smaller than with lower $\Esn$.
For {\tt E10}, the elemental abundance ratio in the recollapsing cloud is consistent with that
from the uniform ejecta ($\Delta \abFe{C} = 0.09$).
However, for {\tt E5}, the iron abundances are enhanced by $\sim 0.4$ dex.
A part of the innermost layers that is initially in the direction to the void
moves to the contact surface between the shell and filament 
because of the turbulence driven by the energetic explosion.
This region contains 40.6\% of the total iron mass that eventually falls back into the central MH.
This results in the enhancement of iron relative to the elemental abundance of the uniform ejecta
($\Delta \abFe{C} = -0.33$),
suggesting that the hydrodynamics in the shell can induce a $\sim 0.3$ dex of variation in the
elemental abundance ratio.

\subsubsection{External enrichment}
\label{sec:EE}

For the EE case, in general, although metals reach a neighboring halo, 
they are hardly mixed with the gas in the central
star-forming core \citep{Chen17, Chiaki18}.
If gas density in the halo already exceeds a threshold value $n_{\rm H,th} \sim 10 \ \percc$, 
the metals only superficially enrich the envelope of the cloud due to the pressure gradient.
The metals can penetrate into the cloud center only if
the SN energy is sufficiently strong \citep{Chen17} or 
if the halo merges with other halos \citep{Chiaki18}. 
For {\tt E10}, metals reach a neighboring halo initially at a distance $D = 418$ pc 
from the central MH.
The halo merges with two other halos (cyan circles in Fig. \ref{fig:snapshots_ini}) 
at $\sim 1$ Myr after the SN explosion and then EE occurs.
At $\trecol = 37.2$ Myr, the merged cloud becomes closer to the central MH with $D = 333$ pc (Fig. \ref{fig:snapshots_E10_EE}).
The radius of the region with densities above $n_{\rm H,th}$, which can trap metals, is
$\Rcl \simeq 20$ pc (Fig. \ref{fig:rn_fin}).

Only the outer layers of the ejecta with initial mass coordinates $\Mr > 2.5 \ \Msun$ reach the 
neighboring halo.
The fraction of the materials incorporated into the cloud is $\ffb (\Mr) \sim 10^{-3}$ at $\Mr > 4 \ \Msun$
(the blue curve in Fig. \ref{fig:mf_ini}).
This fraction can be explained with the simple estimation used in Section \ref{sec:IE} as
$\ffb = \pi \Rcl ^2 / 4 \pi D^2 = 9.01\E{-4}$.
With the mass of the enriched cloud $\Mcl = 1.40\E{5} \ \Msun$, we can estimate the C abundance of the enriched cloud
to be $\abA{C} = 2.30$ from Eq. (\ref{eq:AC}),
consistent with the simulation result $\abA{C} = 2.48$ (Table \ref{tab:metallicity}).
The abundance is much smaller than those in the case with IE by $\sim 2$ dex
because the metals are diluted by the larger mass of pristine gas by two orders of magnitude than
the typical cloud mass ($2000 \ \Msun$).
The cloud mass is much larger because the halo hosting the cloud grows through a set of mergers before EE occurs.

\begin{figure*}
\includegraphics[width=0.7\textwidth]{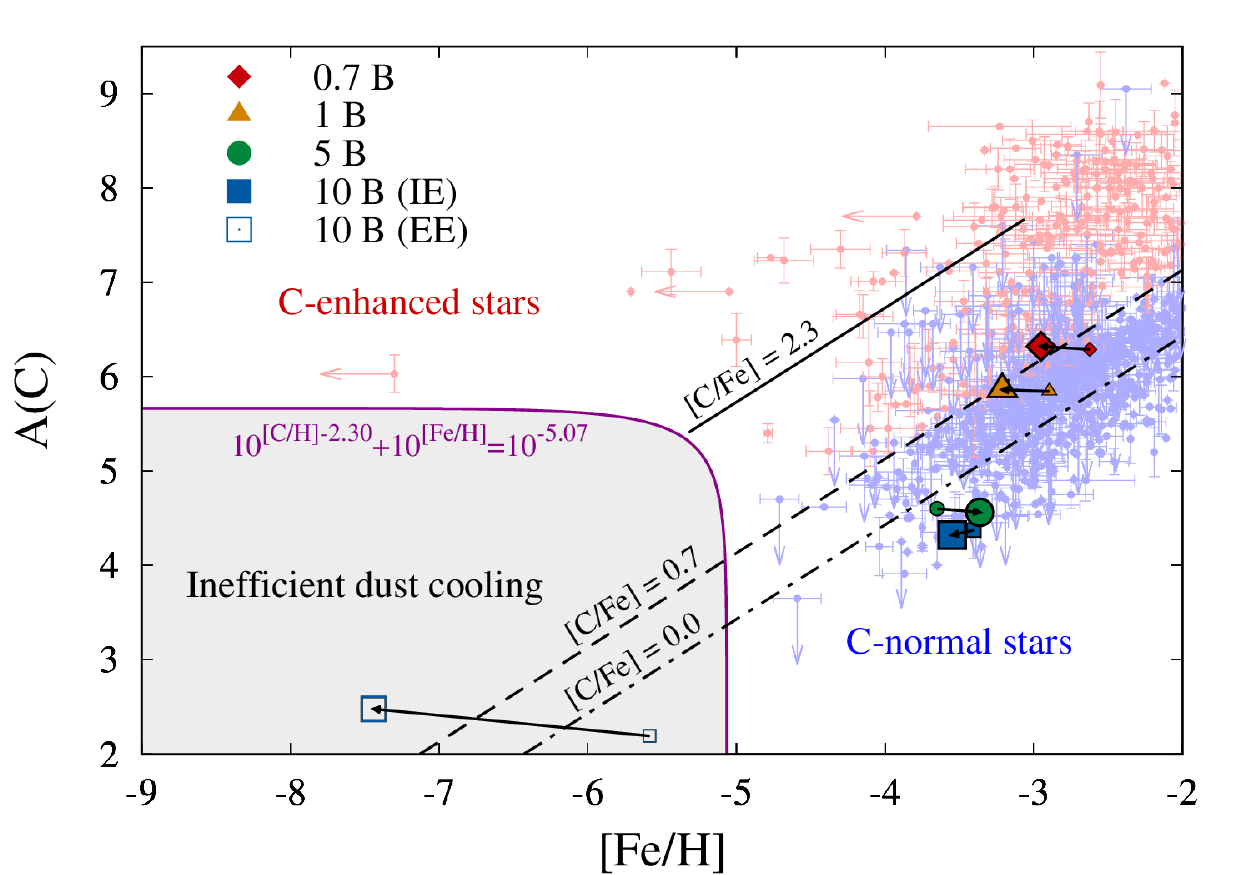}
\caption{
C and Fe abundances of the enriched clouds from the stratified (large symbols) and uniform (small symbols) ejecta
for explosion energies $\Esn = 0.7$ B (red), 1 B (orange), 5 B (green), and 10 B (blue).
The close and open symbols depict the results in the internally and externally enriched clouds, respectively.
The light blue and red dots depict the abundances of observed C-normal and C-enhanced stars, respectively,
taken from SAGA database \citep{Suda08}.
The dot-dashed, dashed, and solid lines show the solar abundance ($\abFe{C} = 0$),
the criterion for CEMP stars \citep[$\abFe{C} = 0.7$;][]{Aoki07}, and 
the division of Group III and Group II stars \citep[$\abFe{C} = 2.3$;][]{Yoon16, Chiaki17}, respectively.
In the grey shaded region below the purple solid curve $10^{\abH{C}-2.30} + 10^{\abH{Fe}} = 10^{-5.07}$, 
clouds collapse stably against fragmentation for the lack of dust cooling \citep{Chiaki17}.
}
\label{fig:Fe_C}
\end{figure*}

%%%%%%%%%%%%%%%%%%%%%%%%%%%%%%%%%%%%%%%%%%%%%
% DISCUSSION %%%%%%%%%%%%%%%%%%%%%%%%%%%%%%%%%%%%%
%%%%%%%%%%%%%%%%%%%%%%%%%%%%%%%%%%%%%%%%%%%%%
\section{Discussion}
\label{sec:discussion}

\subsection{Prediction of observations}
\label{sec:obs}

If low-mass stars ($M_* < 0.8 \ \Msun$) form in the enriched clouds,
they can survive for the Hubble time \citep{Smith15, Chiaki19}.
Then, if some of the stars are accreted onto our Milky Way halo or local dwarf galaxies through
cosmic structure formation, they can be observed as EMP stars
\citep[][and references therein]{Frebel15}.
The colored symbols in Fig. \ref{fig:Fe_C} show the distribution of the stars on the $\abA{C}$-$\abH{Fe}$ plane.
The closed and open symbols depict the abundances of stars forming in the internally and externally enriched
clouds (hereafter called IE-stars and EE-stars), respectively.
The small and large symbols depict the abundances for the uniform and stratified ejecta, respectively.
Interestingly, the Fe abundances are shifted while C abundances hardly change.
The mass of the outer layers is larger than that of the inner layers, where
the fall back fraction fluctuates, and
the mass-weighted average fraction $\langle \ffb \rangle$ is closer to
the fraction $\ffb (\Mr)$ in the outer layers (Fig. \ref{fig:mf_ini}).

For comparison, we also plot the C and Fe abundances of observed C-normal and C-enhanced stars with blue and red dots,
respectively \citep[taken from the SAGA database;][]{Suda08}. 
The stars with below and above $\abFe{C} = 0.7$ (dashed line) are defined as C-normal stars and
CEMP stars, respectively \citep{Aoki07}.
C-normal stars are distributed around the line $\abFe{C} = 0$ (dot-dashed line).
CEMP stars with $\abH{Fe} < -3$ are classified into two sub-groups according to their distributions
\citep{Yoon16}.
Group III stars are distributed around the horizontal line with a constant $\abA{C}$
while Group II stars are distributed along the line with a constant $\abFe{C}$ and 
just above the region where C-normal stars are distributed.
The $\abA{C}$ 
show the large star-to-star scatter of $\sigma (\abA{C}) \sim 1$ dex with a fixed metallicity
for all groups.
This can be considered to reflect the variation of mass and explosion energy of few progenitor stars in the early stage of
galactic chemical evolution \citep{Ryan96, Cayrel04}.

From our simulations, the IE-stars are distributed around the line $\abFe{C} = 0$ (dot-dashed line) with $\sim 1$ dex
deviation.
This is consistent with the distribution of observed C-normal and Group II stars.
For $\Esn \geq 5$ B (green circle and blue square), the IE-stars have low $\abFe{C}$ values 
(bottom right in Fig. \ref{fig:Fe_C}) because $\abFe{C}$ for a uniform ejecta ($-0.18$ and $-0.65$) 
is smaller than the solar abundance ratio, and
is unchanged or depleted by the effect of the stratified ejecta.
For $\Esn \leq 1$ B (red diamond and orange triangle),
$\abFe{C}$ is larger (0.48 and 0.31) for the uniform ejecta and further enhanced 
for the stratified ejecta (see Section \ref{sec:IE} ii).
The stars are distributed around the critical line for CEMP stars $\abFe{C} = 0.7$ (dashed line).
For $\Esn = 0.7$ B, the stars would be classified as CEMP stars (red diamond).
In this work, we fix the progenitor mass to $25 \ \Msun$, but
the range of explosion energies (0.7--10 B) can explain the distribution of observed C-normal 
and CEMP Group II stars with a scatter of $\sim 1$ dex.

For EE-stars, both $\abA{C} = 2.48$ and $\abH{Fe} = -7.44$ are smaller than the
ones of IE-stars by two orders of magnitude because metals are mixed with 
a large mass of pristine gas during a merger of three host halo progenitors which induces
EE (Section \ref{sec:EE}).
As a result, $\abA{C}$ and $\abH{Fe}$ of the EE-stars exist in the low metallicity 
region where stars have not been observed (grey shaded region in Fig. \ref{fig:Fe_C}).
We can consider several reasons why these EE-stars with extremely small metallicities 
have not been observed.
First, because the metallicity for IE is larger than that for EE,
the C-enhanced abundance pattern resulting from EE may be washed out
by the IE of the neighboring cloud itself.
Halos only externally enriched may be so rare that EE-stars have not
so far been observed among the current samples of EMP stars \citep{Hartwig18}.
Second, clouds with insufficient metal content ($\abA{C} \lesssim 6$) should contain
a negligible amount of dust grains, which is the main coolant that can induce cloud fragmentation.
\citet{Chiaki17} estimate the elemental abundances below which dust cooling works inefficiently
as $10^{\abH{C}-2.30} + 10^{\abH{Fe}} = 10^{-5.07}$ (purple curve in Fig. \ref{fig:Fe_C}).
The EE-stars are plotted below this curve, where
clouds can collapse stably against fragmentation, and a single massive star ($ > 10 \ \Msun$) 
is likely to form.
Such massive stars can not be observed because they will not survive for 13.6 Gyr between
the recollapsing redshift $\zrecol = 22.5$ and the present day.

\subsection{Caveats}
In this work, we perform numerical simulations of the transition from Pop III stars to Pop II stars,
considering the non-uniform structure of SN ejecta with realistic nucleosynthesis models.
Because of computational limits, there are several caveats in the simulations.

\subsubsection{Resolution of simulations}
\label{sec:reso}

In the simulations, the SPH particle mass is $m_{\rm p, gas} = 4.45 \ \Msun$,
and the number of neighboring particles is $\Nngb = 64$.
The number of particles that resolve half of the Jeans length $\Rjeans$ of the collapsing clouds with a density 
$\nH = 10^{3} \ \percc$ and temperature $T = 100$ K is
\begin{equation}
\frac{\Rjeans}{h} = 3.91 
\left( \frac{m_{\rm p,gas}}{4.45 \ \Msun} \right)^{-1/3}
\left( \frac{\nH}{10^3 \ \percc} \right)^{-1/6}
\left( \frac{T}{100 \ {\rm K}} \right)^{1/2}
\label{eq:reso}
\end{equation}
where $h = (3 \Nngb m_{\rm p,gas} / 4 \pi \rho)^{1/3}$ is the smoothing length.
This resolution is lower than adaptive mesh refinement (AMR) simulations of chemical enrichment 
\citep{Ritter12, Smith15, Chiaki19}.

The Str\"{o}mgren radius is not properly resolved just before the SN explosion (Section 2.2).
In an AMR simulation of \citet{Ritter12}, the region within a few pc is fully ionized by a Pop III star
with a mass $40 \ \Msun$, and the density of the \HII \ region decreases down to a few $\percc$. 
To resolve the \HII \ region, we could use refinement techniques, such as particle splitting.
Also, de-refinement of the particles are required just before the SN explosion to follow the expansion
of SN shells for the reason stated below.
The refinement and de-refinement procedures would induce numerical errors \citep{ChiakiYoshida15}.
Eitherway, at the time of the SN, we inject metals and thermal energy in a sphere with a radius $\sim 3$ pc
comparable to the expected radius of the H {\sc ii} region.
Therefore, in this work, we do not refine or de-refine particles for the formation of \HII \ region.

\begin{table*}
\begin{minipage}{10.0cm}
\caption{Metallicity in enriched clouds in inner region}
\label{tab:metallicity2}
\begin{tabular}{ccccccc}
\hline
$^{1}\Esn$       & $^{2}$enrich. & $^{3}R_{\rm J}/2$ & $^{4}\abA{C}$ & $^{5}\abH{Fe}$ & $^{6}$[C/Fe]  & $^{7}\Delta$[C/Fe] \\
$[{\rm B}]$      &               & [pc]              &               &                &               &                    \\
\hline \hline                                                                                                              
$ 0.7$           & IE            & $2.04$            & $6.42$        & $-3.05$        & $ 1.04$       & $ 0.56$            \\
$ 1.0$           & IE            & $2.09$            & $5.92$        & $-3.27$        & $ 0.76$       & $ 0.45$            \\
$ 5.0$           & IE            & $1.95$            & $4.64$        & $-3.05$        & $-0.74$       & $-0.56$            \\
$10.0$           & IE            & $1.53$            & $4.13$        & $-3.75$        & $-0.55$       & $ 0.10$            \\
                 & EE            & $2.04$            & $2.24$        & $-7.68$        & $ 1.49$       & $ 2.14$            \\
\hline
\end{tabular}
\medskip \\
Note --- 
(1) Explosion energy.
(2) Enrichment mode.
(3) Half of the Jeans radius.
(4--6) Average carbon and iron abundances and relative abundance ratio $\abFe{C}$ within $\Rjeans / 2$.
(12) Difference of $\abFe{C}$ between the stratified and uniform ejecta.
\end{minipage}
\end{table*}

We cannot use particle splitting after a SN explosion
to resolve a cooling shell and a recollapsing cloud
because particles in the dense regions such as ejecta and shocked shell would be refined
and blown away by SN blastwaves to the outer region filled with coarse particles.
In the interacting regions between particles with different masses, spurious surface tension would affect 
the estimation of hydrodynamic force \citep{Saitoh13}.
Although we barely resolve the Jeans length of the recollapsing clouds,
the elemental abundance ratio for {\tt E1} is consistent with the results of higher-resolution 
simulations of \citet{Ritter15} and \citet{Sluder16} for the same $\Esn$.
Also, the low-resolution simulations enable us to investigate the metal enrichment for the wide range
of explosion energies $\Esn = 0.7$--$10$ B.

\subsubsection{Abundance in inner region of enriched clouds}
\label{sec:mixing_inner}

Fig. \ref{fig:rn_fin} shows that the elemental abundance still has $\sim 1$ dex deviation
at half of the Jeans length $\Rjeans$ from the collapse center (black dashed line).
Further, $\abH{Fe}$ systematically decreases or increases toward the center
within $\Rjeans$ for {\tt E0.7} and {\tt E5}, respectively.
In the succeeding run-away collapse phase, the mass of the cloud core
decreases as the maximum density increases \citep{Larson69}.
Therefore, the stars that finally form might have the elemental abundance of the innermost region of the cloud.
We estimate the average abundance of the cloud within $\Rjeans / 2$ (red dotted lines).
As Table \ref{tab:metallicity2} shows, the difference of abundance ratio $\Delta \abFe{C}$ 
becomes larger ($\pm 0.6 $ dex for IE) 
than that within $\Rjeans$ (Table \ref{tab:metallicity}).
The region within $\Rjeans / 2$ is resolved only two smoothing lengths of SPH particles (Eq. \ref{eq:reso}), 
and thus we can hardly conclude that the abundances would reflect this value.
Simulations with higher resolution are thus required to determine the elemental abundances of 
stars that would eventually form.
We will perform AMR simulations to clarify the metal mixing in
higher density regions in forthcoming papers.

\subsubsection{Mass and multiplicity of Pop III}
\label{sec:multiPopIII}

In this paper, we fix the Pop III stellar mass to $25 \ \Msun$.
This assumption is justified from some numerical simulations showing that
the peak of the Pop III mass distribution is around 
$30 \ \Msun$ \citep{Susa14, Hirano14, Hirano15}.
Also, \citet{Ishigaki18} find that elemental abundances of most EMP stars are best fit with a
Pop III hypernova model with a progenitor mass $25 \ \Msun$.
However, these studies also suggest a range of stellar mass ($\sim 10$--$1000 \ \Msun$),
which can yield a variety of elemental abundance patterns.
We need to study metal enrichment with different Pop III stellar masses to understand the full 
abundance distribution of EMP stars (Fig. \ref{fig:Fe_C}).
Further, we assume that the MH hosts only a single Pop III star. 
Recent numerical simulations show that multiple Pop III stars form through the fragmentation
of the accretion disk \citep{Turk09, Clark11, Greif12, Susa14, Hirano17}.
The elemental abundance of enriched clouds will be the superposition of elemental abundances of
multiple progenitor stars \citep{Feng14, Ritter15}.

We consider a single MH with a mass $3\E{5} \ \Msun$.
Although this is the typical mass of Pop III hosting MHs \citep[Fig. 12 of][]{Chiaki18},
MHs have a wide mass range from $2\E{5} \ \Msun$ to $3\E{6} \ \Msun$.
For low-mass MHs ($\lesssim 10^6 \ \Msun$), a massive Pop III star ($\gtrsim 100 \ \Msun$)
can create the H {\sc ii} region with radii larger than the virial radius, and
the SN shell can expand without losing its thermal and kinetic energy. 
Metals can reach neighboring MHs, and EE mainly occurs.
On the other hand, for Pop III stars with masses $\lesssim 100 \ \Msun$ and explosion energies
$ < 10$ B, IE is the main enrichment channel. 
For high-mass MHs ($\gtrsim 10^6 \ \Msun$), IE mainly occurs
even for massive Pop III stars and energetic SNe \citep{Chiaki18}.

We do not include the effect of streaming velocity,
which in general occurs due to the different evolution of dark matter and baryon 
density fluctuations before the recombination \citep{Tseliakhovich10}.
This results in the delayed star formation caused by the offset of dense gas clumps from 
dark matter potential wells \citep{Chiou18, Chiou19, Druschke19}.
Star formation occurs in more massive halos ($\sim 10^7 \ \Msun$) than zero-streaming
velocity cases, where EE hardly occur even for high $\Esn$ \citep{Chiaki18}.
Previous studies also predict that clouds with surpersonic turbulence may fragment 
to massive Pop III star clusters \citep{Hirano18}, indicating that the
mixing of metals from multiple sources may be more significant.

\subsubsection{Stars with peculiar elemental abundance}
\label{sec:CEMP}

We use SN models that reproduce the elemental abundances of C-normal EMP stars
with $-0.65 < \abFe{C} _0 < 0.48$.
However, $\sim 80$\% of EMP stars with metallicities $\abH{Fe} < -4$ show C-enhanced abundance
pattern \citep{Yoon18, Norris19}. 
Several scenarios have been proposed to explain the CEMP star formation.
One is the intrinsic enrichment scenario in which parent clouds of the stars are
enriched by C-enhanced interstellar gas.
The source of C-rich gas is considered to be SNe with a large fallback
of Fe-rich innermost layers into the central remnant
and ejection of C-rich outer layers into the ISM \citep[faint SNe;][]{Umeda03}.
The other is the extrinsic enrichment scenario in which C-rich gas accretes onto
the surface of stars from their binary companion in the asymptotic giant branch (AGB) phase \citep{Suda04, Komiya20}.

We can simply apply this work to the faint SN model by assigning the corresponding radial distribution 
of elemental mass fraction to the metal tracer particles.
As Fig. \ref{fig:mf_ini} shows, the fraction of metals $\ffb (\Mr)$ incorporated into the enriched
ejecta is almost constant in the outer layers
but deviated from this constant value in the innermost layers.
Since elements lighter than Fe are dominant in the
innermost region of faint SN ejecta, the slight deviation of their abundances might be
seen for faint SNe.

Stellar rotation can also modify the elemental abundance in the final phase of stellar evolution.
The elemental abundance of materials blown away from the ensuing SN explosion shows C, N,
and s-process element enhancement \citep{Meynet06, Choplin17}.
In addition, the rotating stars are considered to explode as jet-like SNe.
This model is introduced to explain Si-deficient of the star HE $1424-0241$ and 
Zn-enhancement of the star HE $1327-2326$ \citep{Tominaga09, Ezzeddine19}.
The elemental abundance of clouds enriched from this type of SNe should strongly depend on
the direction of the jets against the three-dimensional structure of the intergalactic medium.
We will see the enrichment process of jet-like SNe in forthcoming papers.

%%%%%%%%%%%%%%%%%%%%%%%%%%%%%%%%%%%%%%%%%%%%%
% CONCLUSION %%%%%%%%%%%%%%%%%%%%%%%%%%%%%%%%%%%%%
%%%%%%%%%%%%%%%%%%%%%%%%%%%%%%%%%%%%%%%%%%%%%
\section{Conclusion}
\label{sec:conclusion}

We perform numerical simulations focusing on the metal enrichment from Pop III SNe, 
considering a stratified structure of ejecta.
Here we consider normal core-collapse supernova (CCSN) models with $-0.65 < \abFe{C} _0 < 0.48$.
We find that SN shells fall back into the central minihalo in all models.
The abundance ratio $\abFe{M}$ in the recollapsing clouds
deviates from that from the uniform ejecta by at most $\pm 0.4$ dex for any element M.
Overall, the fraction $\ffb (\Mr)$ of metals falling back into the recollapsing clouds is almost
constant regardless of the initial mass coordinate $\Mr$ of the ejecta.
The slight deviation from the average abundance of the ejecta is mainly from the turbulent motion
of the hot innermost layers.
The metallicity range of these clouds is $-3.55 < \abH{Fe} < -2.95$ and resulting
C to Fe abundance ratio is $-0.56 < \abFe{C} < 0.84$.
If the stars directly inherit the elemental abundance of their host clouds,
the abundances of the stars are consistent with those of the observed C-normal and Group II CEMP stars.

In addition, for the largest explosion energy (10 B), a neighboring halo is also enriched.
Only the outer layers rich with Ca or lighter elements reach the halo, where
the abundance ratio of C to Fe is $\abFe{C} = 1.49$.
This means that C-enhanced metal-poor (CEMP) stars can form from the CCSN 
with the average abundances ratio $\abFe{C}_0 = -0.65$.
However, the metallicity of this cloud $\abH{Fe} = -7.44$ is
smaller than in the IE cases by two orders of magnitude because
the halo contains a large mass of pristine gas ($\sim 10^5 \ \Msun$) before it is enriched.
In the low-metallicity region, no low-mass stars have so far been observed.
With a statistic analysis, the hypothesis that these stars escape 
the detection is ruled out at a 99.9\% confidence level with $\sim 500$ samples of EMP stars observed
so far \citep{Magg19}.
An explanation of the non-detection is that
such a low-metallicity cloud will collapse stably without fragmentation because of
a lack of cooling from dust grains.
Massive stars are likely to form and not be observed due to their short lifetime
compared to the Hubble time \citep{Chiaki17}.

In this paper, we have studied the metal enrichment from normal CCSNe with a fixed progenitor mass.
Nowadays $\sim 10^6$ stars have been observed with large survey campaigns, and 
$\sim 5000$ EMP stars have been identified with follow-up spectroscopic measurements.
These observations have revealed that there are various classifications of 
EMP stars with an enhancement or depletion of specific elements.
We will extend our numerical methods to the other types of SN models
such as rotating stars, faint SN models and jet-like SN models (Section \ref{sec:CEMP})
to construct the comprehensive formation models of statistical samples of EMP stars.
These studies will uncover the general process of chemical enrichment in the ISM
and the origins of elements that compose the Universe.

%%%%%%%%%%%%%%%%%%%%%%%%%%%%%%%%%%%%%%%%%%%%%%
% ACKNOULEDGEMENTS %%%%%%%%%%%%%%%%%%%%%%%%%%%%%%%%
%%%%%%%%%%%%%%%%%%%%%%%%%%%%%%%%%%%%%%%%%%%%%%
\section*{ACKNOWLEDGMENTS}

We thank John H. Wise
for fruitful discussions and comments.
GC is supported by Overseas Research
Fellowships of the Japan Society for the Promotion of Science (JSPS)
for Young Scientists.  
The numerical simulations and analyses in this work are carried out on
XC40 in Yukawa Institute of Theoretical Physics (Kyoto University),
Comet in SDSC, Stampede2 in TACC on NSF's XSEDE allocation AST-120046, and
PACE/HIVE clusters in the Georgia Institute of Technology.
The figures in this paper are constructed with the
plotting library {\sc gnuplot} and {\sc matplotlib} \citep{matplotlib}.

%%%%%%%%%%%%%%%%%%%%%%%%%%%%%%%%%%%%%%%%%%%%%%
% DATA AVAILABILITY %%%%%%%%%%%%%%%%%%%%%%%%%%%%%%%
%%%%%%%%%%%%%%%%%%%%%%%%%%%%%%%%%%%%%%%%%%%%%%
\section*{Data availability}

The data underlying this article will be shared on reasonable request 
to the authors.

%%%%%%%%%%%%%%%%%%%%%%%%%%%%%%%%%%%%%%%%%%%%%%
% APPENDIX  %%%%%%%%%%%%%%%%%%%%%%%%%%%%%%%%%%%%%%
%%%%%%%%%%%%%%%%%%%%%%%%%%%%%%%%%%%%%%%%%%%%%%

%\appendix
%\section{Dust cooling rates with grain growth}

%%%%%%%%%%%%%%%%%%%%%%%%%%%%%%%%%%%%%%%%%%%%%
% REFERENCES %%%%%%%%%%%%%%%%%%%%%%%%%%%%%%%%%%%%
%%%%%%%%%%%%%%%%%%%%%%%%%%%%%%%%%%%%%%%%%%%%%

\label{lastpage}

\end{document}